\documentclass[12pt,preprint]{aastex}
\shortauthors{Subrahmanyan, Hunstead, Cox, McIntyre}
\shorttitle{Giant radio galaxy J0515$-$8100}
\begin{document}

\title{SGRS J0515$-$8100: a Fat-Double giant radio galaxy}

\author{Ravi~Subrahmanyan,\altaffilmark{1} R.~W.~Hunstead,\altaffilmark{2} 
N.~L.~J.~Cox,\altaffilmark{2,3,4} and V.~McIntyre\altaffilmark{1}}

\altaffiltext{1}{Australia Telescope National Facility, CSIRO, 
P O Box 76, Epping, NSW 1710, Australia.}
\altaffiltext{2}{School of Physics, University of Sydney, NSW 2006, Australia.}
\altaffiltext{3}{Faculty of Physics and Astronomy, University of Utrecht, P. O. Box 80000, 
NL-3508 TA Utrecht, The Netherlands}
\altaffiltext{4}{Presently at Astronomical Institute ``Anton Pannekoek'', 
University of Amsterdam, NL-1098 SJ Amsterdam, The Netherlands}

\begin{abstract}
We present here the first detailed study of a giant radio galaxy of the Fat-Double type.  
The lobes of the double radio galaxy SGRS J0515$-$8100 have transverse widths that are 1.3 times 
their extent from the center, their surface brightness is the lowest 
among known giant radio sources and the lobes have relatively steep radio spectra. 
We infer that these wide lobes 
were created as a result of a highly variable and intermittent jet whose
axis direction also varied significantly: the Fat-Double lobes in this giant radio source
are a result of the ejection and deposition of 
synchrotron plasma over a wide range of angles over time rather than the expansion
of relic lobes.  Additionally, the optical host shows evidence for an ongoing
galaxy-galaxy interaction. SGRS J0515$-$8100 supports the hypothesis
that interactions with companions might perturb the
inner accretion disk that produces and sustains the jets at the centers of
active galactic nuclei. As a result, it appears unnecessary to invoke black-hole coalescence
to explain such morphologies, implying that the corresponding event rates predicted for 
gravitational wave detectors may be overestimates.
\end{abstract}


\section{Introduction}

The interaction between jets of relativistic plasma---that are generated in
inner accretion disks of active galactic nuclei---and the ambient gaseous
environment creates the synchrotron emitting lobes of powerful radio sources.
The giant radio sources are possibly
powered by central engines with the longest lifetimes; therefore, they are  
potentially probes of the long time history of the nuclear jet stability.
With linear size exceeding 1~Mpc, the giant radio sources have sound crossing times
exceeding 6~Myr; however, spectral aging arguments lead to
radiative ages of order 0.1~Gyr and dynamical arguments suggest ages that are
an order of magnitude larger. Post starburst stellar populations in radio galaxies have
ages 0.5--2.5~Gyr \citep{Ta05} consistent with dynamical ages.  
If the jets vary in power or direction, or if they 
undergo interruptions over these long timescales, 
the structures of giant radio galaxies might be expected to show evidence.
Not surprisingly, the morphologies of giant radio sources do indeed show 
evidence for temporal variations in jet power \citep{Su96}.  
More recently, detailed radio imaging of giant radio galaxies have yielded
some spectacular examples where new jets created in a new epoch of
central engine activity are observed to be ploughing
through relatively relaxed lobes that were presumably deposited in the past (see,
for example, \citet{Sa02} and \citet{Sa03}). The cause for interruptions 
to the jet might be instabilities in the jet
production mechanism, instabilities in the accretion disk, or interruptions
to the fuelling of the central engine. 

The examples of recurrent activity in giants studied to date have been almost exclusively
cases where the jet axis has remained essentially unaltered in the multiple
activity phases.  S- and X-shaped giant radio sources---which are usually interpreted as
cases where the jet axis has varied significantly over time---are extremely rare, consistent 
with the hypothesis that long timescale 
stability in the axis might be necessary for the creation of giant Mpc-size
radio sources \citep{Su96}. In this model, 
giant radio galaxies would not be expected to manifest Fat-Double structure, unless
the wide lobes result from transverse expansion in the relic phase 
after the jets from the central engine switched off.

The evolution of the synchrotron lobes of radio galaxies after the jets cease has
important implications for our understanding of their dynamical interaction with the
ambient medium.  Constraints on models for the dynamical evolution of the relics,
together with statistics of the occurrence of relics in complete samples of radio sources,
translate into constraints on dynamical ages of active radio sources and, therefore, on
the stability and timescale of the nuclear activity and associated fuelling.
Additionally, understanding the evolution of relic giant radio sources has implications for the
physical properties of the intergalactic medium (IGM) that is not directly observable with
the limited sensitivity of present day X-ray telescopes.  The activity lifetimes and the
disappearance rate of extended extragalactic sources are the parameters defining the injection
rate for relativistic plasma and magnetic fields into the IGM.  
Less than 3\% of all double radio sources are relics in existing surveys 
\citep{Gi88} and their rarity is an enigma that is a reflection of our
lack of understanding of the end stages of the radio source phenomenon and how 
radio sources exit the observable parameter space.

In the Sydney University Molonglo Sky Survey (SUMSS; \citet{Bo99}), \hfill\break
SGRS~J0515$-$8100 appeared 
to be a close pair of low-surface-brightness emission
regions that were 3--5 arcmin in size and with their centers separated by 
about $5\arcmin$. Followup observations indicated that the source
is a giant radio galaxy with an unusual Fat-Double structure: the source was included in the
compilation of southern giant radio sources made by \citet{Sa05}.
SGRS~J0515$-$8100 appears to be a rare example of a 
giant radio galaxy in which there has not only been
significant variations in jet continuity but also significant variations 
in the direction of the jet axis.  Additionally, the optical host displays
evidence of an ongoing interaction/merger.  Because this source is a rare type with implications
for the stability in the nuclear jets, formation of giant radio galaxies and for the 
end stages of the evolution of relic lobes in an IGM environment, 
we present here a case study of SGRS~J0515$-$8100.  A journal of the followup observations is
in Table~1.

\section{Radio observations}

The Molonglo Observatory Synthesis Telescope (MOST) was used to make
a 36-cm wavelength image of a $23\arcmin ({\rm RA}) 
\times 23\arcmin {\rm cosec}\delta ({\rm DEC})$  field centered on SGRS~J0515$-$8100.  
Contours of this radio image made with a beam of full width at half 
maximum (FWHM) $43\farcs5 \times 43\farcs0$ at position angle (P.A.) 
$0\degr$ are shown in Fig.~1 overlaid on a SuperCOSMOS digitization of a
UKST blue optical image. The radio image has an rms noise of 0.5~mJy~beam$^{-1}$.  

The radio source was subsequently imaged with the Australia Telescope Compact Array (ATCA)
at 12 and 22~cm wavelengths from visibility data obtained in the
210 and 375~m configurations, as well as the longer 1.5A and 6C arrays that
provide baselines up to 6~km.   The flux density scale was set using observations of
PKS~B1934$-$638 whose flux density was adopted to be 11.1 and 14.9~Jy respectively at
12 and 22~cm.  The interferometer data were calibrated and imaged using MIRIAD: 
multi-channel continuum data were imaged using standard bandwidth-synthesis techniques. 
The images were first partially deconvolved using the Clark algorithm \citep{Cl80} to 
remove the effects of unresolved components and then the low-surface-brightness
extended emission was deconvolved using the Steer-Dewdney-Ito algorithm \citep{St84}.
The 22-cm image of the source, made with a beam of $25\arcsec$ FWHM, is shown in
Fig.~2; the rms noise in the image is 0.05 mJy~beam$^{-1}$. The 12-cm image of the source
made with the same beam is shown in Fig.~3; here the rms noise is 0.08 mJy~beam$^{-1}$. 

The radio images show a pair of disjoint radio lobes with 
no evidence for any connecting bridge emission.  Both lobes are wider in their transverse
extents as compared to their axial lengths.  The optical host for the
double radio source is likely to be the $b_{J}$ = 17.2 galaxy that appears at a
location close to the center of the radio source on the sky and at the northern end of the 
southern lobe.  

When the radio source was reconstructed using the longer 1.5A and 6C array visibilities,
and a higher resolution image with beam FWHM $8\farcs6 \times 4\farcs9$  
at P.A. $-34\degr$ was made
at 12~cm wavelength, two compact unresolved components appear within the sky area spanned 
by the double radio source.  The first, at RA: 05$^{h}$15$^{m}$55\fs1,
DEC: $-$80\degr59\arcmin43\farcs0 (J2000), is coincident with the
candidate host galaxy and is presumably the radio core component of the double radio
source.  This core component appears as an unresolved source in both the 12 and 22~cm
images and has a flat spectral index $\alpha \approx 0.1$ between these wavelengths
(we adopt the definition $S_{\nu} \propto \nu^{\alpha}$ for the spectral index, where 
$S_{\nu}$ is the flux density at observing frequency $\nu$). A second unresolved
component is observed at RA: 05$^{h}$15$^{m}$04\fs9, 
DEC: $-$80\degr57\arcmin40\farcs1 (J2000) and appears on the sky
within the northern lobe.  This component has no associated intermediate resolution extended
emission and is, therefore, unlikely to be a hotspot. This
source has a radio spectral index $\alpha \approx -1.05$, has no optical
counterpart on the SuperCOSMOS digital sky survey and is presumably 
an unrelated background object. Both unresolved compact components are estimated to have
angular sizes less than 4\arcsec.   

The southern lobe of the double radio source SGRS~J0515$-$8100 has a transverse width that increases 
to the south and away from the core.  This lobe is bounded at its southern end by a 
relatively bright rim of emission that is discontinuous and concave outwards.
The surface brightness in this lobe increases away from the core and towards this rim.
The disjoint northern lobe is composed of two extended components that
are aligned perpendicular to the source axis: a symmetric
component to the west that has a relaxed appearance 
and a smaller edge-brightened component to the NE; these two components 
are connected by a bridge of emission.  

\subsection{Radio spectral energy distribution}

The radio flux densities of the source components from our MOST and ATCA observations
are listed in Table~2. It may be noted here that
the northern and southern lobes of SGRS~J0515$-$8100 have previously 
been detected in the Parkes-MIT-NRAO (PMN) 
survey \citep{Gr93} at 6~cm wavelength with flux densities of about 13 and 35~mJy respectively; these
values are also listed in the Table.
The northern and southern lobes have straight radio spectra with
steep spectral indices of $\alpha \approx -1.0$  between 36 and 6~cm wavelengths; 
within the errors in the measurements there is
no evidence for any spectral curvature or breaks in the integrated 
spectrum over this wavelength range.  

The distribution in the spectral index over the source was computed using images at
36 and 12~cm wavelength that had been convolved to a common beam of $1\arcmin$ FWHM; 
this is shown in Fig.~4. Only pixels exceeding four times the rms noise in the individual
images were used in forming the spectral index distribution.
Overall the NW part of the northern lobe and the SW part of the southern lobe 
have the flattest spectra.  Excluding the compact core, the extended emission is observed to have
a progressive steepening in the spectral index 
from the ends of the source towards the center.  Additionally, the western parts of the lobes
are relatively flatter in their indices as compared to the eastern parts. 
The bridge connecting the NE and NW components of the northern lobe has
a relatively steeper spectral index.

The $1\arcmin$ FWHM images at 36, 22 and 12~cm have also been used to make a color-color plot,
shown in Fig.~5. At each image pixel ($16\arcsec$ apart) 
that had significant flux density at all three wavelengths, the low frequency spectral index 
$\alpha^{36}_{22}$ (between 36 and 22~cm) is plotted against the high frequency
spectral index $\alpha^{22}_{12}$ (between 22 and 12~cm).  
Each point represents the location of an image pixel
in this plane.  Pixels over the core component have been omitted.  

In the color-color plot, the relatively bright rim along the outer edge of the southern 
lobe, and the NW end of the northern lobe, occupy an elongated concentration running
roughly parallel to and just above the $\alpha^{36}_{22} = \alpha^{22}_{12}$ line.  In these 
regions $\alpha^{36}_{22}$ is in the range $-0.9$ to $-1.3$ and the spectral index  between
22 and 12~cm is steeper than that between 36 and 22~cm by about 0.3.  Along the southern rim, 
the spectral index steepens from west to east.  In the lower surface brightness regions towards the core 
and away from the ends of the two lobes, the image pixels have a wider distribution in color-color space.  
Whereas $\alpha^{36}_{22}$ is between $-1$ and $-1.5$, $\alpha^{22}_{12}$ steepens and takes on values in the
range $-1.5$ to $-2.5$.  The NE component of the northern lobe displays the most
spectral curvature: it occupies a region where $\alpha^{36}_{22} \approx -1$ 
and $\alpha^{22}_{12}$ steepens from $-1.8$ at the outer end to $-2.3$ at its inner end.  The few points
scattered to the right of the $\alpha^{36}_{22} = \alpha^{22}_{12}$ line are from
pixels close to the edges of the emission and probably arise from errors in the 12~cm image.

The parts of the extended emission most distant 
from the core have relatively flatter spectral indices and 
weak spectral curvature in the observed wavelength range.  
In both lobes the spectral index of the extended emission steepens and displays enhanced spectral
curvature towards the center of the double radio source.
The lowest surface brightness regions of the lobes, which are closest to the core,
also have the largest spread in the color-color plane. 
The southern lobe has flatter
indices in the SW regions that have the highest surface brightness; in contrast, the northern lobe
has a spectral gradient across the lobe---the spectrum is flattest at the NW end and progressively
steepens towards the core---although in total intensity the lobe appears symmetric and relaxed.

\subsection{Radio polarization properties}

Images in Stokes Q and U were made at 12 and 22~cm with beam $25\arcsec$ FWHM 
and these were used to
construct images of the distribution in polarized intensity and the orientation of the
observed E-field vectors across the source.
The distribution in the intensity of the linearly polarized emission at 22~cm is shown 
in Fig.~6 using contours overlaid on a grey scale representation of the total intensity image. 
In the southern lobe, the polarized emission is sharply bounded along a straight edge
running roughly NS; the edge is not aligned with the core and 
there is no corresponding feature in the total intensity image along this break.
To the east of the edge, the southern lobe is 40--60\% polarized whereas to the
west, the lobe is 10--30\% polarized.  In the northern lobe, the polarized intensity
as well as the fractional polarization are high at the NW end and along the edge-brightened
boundary of the NE component: here the fractional polarization rises to 50--60\%.

The rotation in the projected E-field vectors
between 12 and 22~cm is consistent with a uniform rotation measure (RM) 
of value +50~rad~m$^{-2}$ over the entire source.  The observed RM is consistent with the
measurements of \citet{Si81} towards other extragalactic sources in this sky region indicating 
that the RM is probably Galactic in origin.  
The distribution in the 
orientation of the projected E-field, corrected for this constant RM, is shown in Fig.~7
overlaid on a grey-scale representation of the polarized intensity and contours of
the total intensity.  The projected magnetic field (assumed to be 
orthogonal to the observed E-field) appears to be oriented circumferentially
along the boundaries of the lobes.  Within the southern lobe and to the east of the
sharp discontinuity in polarized intensity, the magnetic field has large scale order and is 
aligned parallel to the southern bounding rim.  To the west of the NS break the orientation
of the magnetic field, in the regions where it has been reliably detected,  
changes to a roughly NS orientation.

Using images made with FWHM $1\arcmin$ and examining the regions where the Stokes I, Q and 
U flux densities exceeded 3 times the image rms noise, the fractional polarization 
at 12~cm was observed to be higher than that at 22~cm: the
depolarization ratio (DR; computed as the ratio of the percentage polarization at 22~cm to that at
12~cm) has a mean of 0.77 and a small spread with standard deviation 0.17.

\section{Optical observations}

The double-beam spectrograph on the ANU 2.3-m telescope at Siding Spring Observatory 
was used to get a spectrum
of the host galaxy over the wavelength range 6200--7464~\AA; 
this is shown in Fig.~8.  H$\alpha$, [N~{\sc ii}] and [S~{\sc ii}]
were detected in emission and Na{\sc i}D in absorption.  The redshift of the host was estimated 
from the absorption line to be $z=0.1052 \pm 0.0002$; the emission lines have a somewhat 
higher redshift and probably arise in infalling gas, with a relative velocity of about 
210~km~s$^{-1}$, in the foreground of the host.  
However, the [N~{\sc ii}]/H$\alpha$ and [S~{\sc ii}]/H$\alpha$ ratios are as expected for
an active narrow-line radio galaxy and unlike those observed in starburst galaxies indicating 
that the emission lines are from a nuclear narrow-line region.

$V$ and $R$ band images of the field of the host galaxy were made using the imager on the 
ANU 2.3-m telescope; the seeing was about $2\arcsec$ during the observations.  
The $R$-band optical field is shown in Fig.~9: in the panel on the left
we have marked the nearby objects and in the panel on the right the grey scale has been chosen to 
display the tidal tails on the two sides of object `e'.  Fits to objects `a', `c' and `d' show
that their FWHM sizes are within the range found for objects that are clearly stars in the field.
The photometry of the objects in the field is given in Table~3 where the $V-R$ color in the 
standard Johnson-Morgan/Cousins bands was derived 
from the observed $v$ and $r$ magnitudes using the relation
\begin{equation} 
V - R = 1.0073 [ (v - r) + 0.0208 ].
\end{equation}
Objects `a', `c' and `d' are classified as K stars on the basis 
of their $V-R$ colors. The host of the radio source, marked as object `b', and the companion
object with the tidal tails, marked `e', have identical $V-R$ colors and may be classified as
$z=0.1$ E galaxies on the basis of their colors.  In the standard Johnson-Morgan/Cousins
visual band, the apparent magnitude of the host `b' is $m_{V}=16.08$ and that of the 
companion `e' is 18.75.  We infer the absolute magnitude of the host galaxy to be
$M_{V}=-22.4$ and that of its companion to be $-19.7$.  

Spectra of the host galaxy and its companion were made in the optical blue band using 
the ANU 2.3-m spectrograph; the spectra covering the wavelength range 
4000--5600~\AA~ are shown in Fig.~10. The companion is blue shifted relative to the host and
a joint fit to the Ca~{\sc ii} H and K absorption lines gives the line of sight 
velocity difference to be $370 \pm 70$~km~s$^{-1}$.  The connecting tidal stream, together
with the spectra, suggest a dynamical interaction between the host galaxy `b' and the 
companion `e'.

\section{The phenomenology of SGRS~J0515$-$8100}

Adopting a flat adiabatic $\Lambda$CDM cosmology with matter density $\Omega_{m} = 0.3$
and Hubble constant 70~km~s$^{-1}$~Mpc$^{-1}$,  the source is at a luminosity distance
of 486~Mpc and images have a linear scale of 116~kpc~arcmin$^{-1}$.  

The source has a total radio power of $4.8 \times 10^{24}$~W~Hz$^{-1}$ at 1.4~GHz. 
The host has $M_{R}=-23.1$ in the Cousins $R$ band and for this absolute magnitude the
radio power is about a factor 7 below the Fanaroff and Riley (FR) class I/II division 
for radio sources \citep{Fa74,Le96}.
Moreover, the radio power is well below values typical of sources 
of the Fat-Double class \citep{Ow89}.
The projected linear size of the radio source is about 1.04~Mpc and the lobes have 
transverse widths of 0.64~Mpc. SGRS~J0515$-$8100 is a giant radio galaxy and
the axial ratio, computed as the ratio of the maximum linear size
to the width, is about 1.6.

At 22~cm wavelength, most parts of the lobes of SGRS~J0515$-$8100 have
surface brightness about 10~mJy~arcmin$^{-2}$.   
Assuming standard minimum energy assumptions \citep{Mi80}, we estimate that 
the relativistic plasma in the lobes of SGRS~J0515$-$8100 has energy 
density $6 \times 10^{-15}$~J~m$^{-3}$. 
The source has the lowest surface brightness among all double
radio sources that we know, and is just a factor of 3 above the surface brightness
of the prototypical cluster-wide halo source in the Coma cluster.
When compared to giant double radio sources
discovered in the WENSS \citep{Sc00} and the VLA NVSS
\citep{Ma01}, SGRS~J0515$-$8100 has the lowest lobe synchrotron energy density. 
The WENSS, NVSS and SUMSS are wide sky area surveys that reach 5-$\sigma$ 
surface brightness detection limits equivalent to 3--6~mJy~arcmin$^{-2}$ at
22~cm wavelength and, therefore, SGRS~J0515$-$8100 is probably one of the lowest surface brightness
extended sources we might expect to have detected in surveys to date.

\subsection{SGRS~J0515$-$8100: a Fat-Double relic giant radio galaxy?}

Powerful radio galaxies, including giant radio 
galaxies, have relatively large axial ratios of about 5--5.6 \citep{Su96}. 
Double radio galaxies of the Fat-Double class \citep{Ow89} have smaller axial ratios.
The prototypical Fat-Double is 3C310; other examples in the literature are 3C386,
3C314.1, Fornax~A and possibly B2 0924+30; these sources have axial ratios in the range 1.7--2.4.
The Fat-Doubles have
diffuse lobes that appear relaxed and without hotspots or any compact features, the surface
brightness in the lobes diminishes towards their edges and most
have lobe spectral indices $\alpha \approx -1$.  Although core radio emission has been 
detected in the hosts of many of these sources, it is believed that
jets from the central engine have stopped feeding the lobes and that
their lobes are relics of past activity. 
Fat-Double sources lie near the FR I/II break in radio power and this is consistent
with the hypothesis that they were
FR~II sources whose luminosity dropped to the present value while their lobes expanded
and evolved to their current relaxed state.

The low surface brightness, low energy density, lack of hotspots and steep spectral index 
are suggestive of an interpretation in which the lobes of SGRS~J0515$-$8100 
are relics of past activity and have attained their present state as a result of 
the disappearance of hotspots and any bright structures owing to relaxation
within the lobes accompanied by expansion losses and synchrotron aging.
With its relatively small axial ratio, intermediate radio power and fairly relaxed appearance, 
SGRS~J0515$-$8100 might be the first example of a giant radio galaxy in the Fat-Double class.
Moreover, the high fractional polarization observed within the lobes of SGRS~J0515$-$8100 
and the circumferential B field orientation along the edges of this source are characteristic
of the class of Fat-Doubles.  

However, SGRS~J0515$-$8100 has features that are inconsistent with relic Fat-Doubles.  
The source has a significant emission gap between the 
two lobes and such gaps have not been observed in Fat-Doubles in the literature. 
The northern lobe appears to be composed of a symmetric and diffuse structure to the west 
which drops off radially in surface brightness as might be expected for a relaxed relic lobe; 
however, the northern lobe has a NE extension that has a brightening along its outer edge
and this component does not have a relaxed structure. The southern lobe is also edge brightened.  
The absence of hotspots in SGRS~J0515$-$8100 suggests that currently there are no jets
from the central engine injecting energy into the lobes.  However, 
the presence of edge brightening in the NE and southern lobes suggests that energy was being 
injected at the ends of these lobes in the recent past, likely within a time 
corresponding to the sound crossing time across
the lobes.  This suggests that the wide lobes are not a result of expansion in a relic phase
after the jets ceased and the lobes relaxed internally.

\subsection{Dynamical evolution}

Significant depolarization has been observed in regions where the
fractional polarization is high and where the field is uniform: the
DR is unlikely due to beam depolarization.  A significant part of
the observed uniform Faraday rotation of +50~rad~m$^{-2}$ is clearly external to the source and
presumably originates in the Galaxy;  however, a part
of the observed RM might arise from entrained plasma that causes internal Faraday
rotation and depolarization.  If the magnetic field 
has an energy density equal to the energy density of the relativistic plasma in the lobes, 
which was estimated assuming standard minimum energy assumptions, a uniform 0.08~nT field 
threads a large part of the lobes.  The observed DR would arise if thermal material with 
particle number density $10^{2}$~m$^{-3}$ is entrained.  
In this case, the generalized sound speed in the lobes
would be as low as 0.0003$c$ and the sound crossing time within the bright structures
along the southern end of the source could be as large as 1~Gyr.  

The host of SGRS~J0515$-$8100 is not in any rich group or cluster environment;  
it is a field galaxy with a companion and is 
presumably located in the filamentary large scale structure of the present epoch.
The Mpc-scale radio structure of this giant radio galaxy is outside any 
coronal gaseous halo that might be associated with the host galaxy and is expected to
be embedded in the warm-hot intergalactic medium (WHIM) that traces large scale filaments.
This WHIM constitutes about 30--40\% of the total baryons and 
is expected to have a temperature 0.01--0.5~keV within the filaments that represent
overdensities in the range 5--200 \citep{Ce99,Da01}.  
The thermal gas pressure in these unvirialized regions is at most $10^{-15}$~Pa. 
Today, the non-thermal gas pressure in the radio lobes of SGRS~J0515$-$8100 is inferred to
be $2 \times 10^{-15}$~Pa, which is just above the range for the external pressure;
therefore, during the evolutionary history the lobes are expected to have been
overpressured with respect to the ambient gas.
The expansion speed of the lobes of SGRS~J0515$-$8100 is limited by the inertial mass of
the WHIM thermal gas, which has a density in the range 0.7--$30 \times 10^{-27}$~kg~m$^{-3}$.
Ram pressure balance indicates that lobe expansion speed $v \la 0.001$--$0.005c$.

The lobes of the radio source are likely to contain thermal gas as a result of entrainment 
at the termination shock at the ends of the source where the jets meet the IGM.  Additionally,
entrainment may occur during the propagation of the jets out of the environment of the
host galaxy.  The gaseous environment is likely to be the WHIM discussed above 
and, even if a significant fraction of the IGM is
entrained, the sound speeds within the lobe plasma could be small enough so that the lobes
might expand and be ram pressure confined at speeds $v \la 0.001$--$0.005c$.  
In this case, lobe expansion by a factor of 2
would require 0.25--1.3~Gyr.  However, if the density of the entrained thermal gas in the
lobes exceeds about $20$~m$^{-3}$, and takes on values closer to 
$10^{2}$~m$^{-3}$ as suggested by the DR measurements, such an entrainment is likely
a result of interactions with relatively dense gas in the interstellar medium of the
host galaxy and its immediate environment.  High density entrainment would limit 
lobe expansion speeds to well below 0.001$c$ and expansion by a factor of 2 would require 4~Gyr.

\subsection{Spectral evolution}

The energy density in the lobes corresponds to an equipartition magnetic field of 0.08~nT, 
which is a factor of 5 less than the magnetic field equivalent to the energy density in the
cosmic microwave background radiation (CMBR).  Assuming that the lobes have a tangled magnetic
field and that they have not expanded adiabatically by more than a factor of 2 since the
jets stopped feeding the lobes, spectral aging in the lobe non-thermal gas would have
been dominated by inverse-Compton losses against the CMBR. 

Edge-brightened double radio sources in which jets are observed to be powering the
lobes usually have spectral indices $\alpha \approx -0.5$ at the hotspots and we assume that
during the active phase the injection index in SGRS~J0515$-$8100 was $\alpha \approx -0.5$.  
During the active phase, simple models for the spectral aging of the relativistic
plasma lead to a spectral break steepening the index to $\alpha \approx -1$ 
at higher frequencies.  The color-color plot in Fig.~5 shows that all parts of the source 
have $\alpha^{36}_{22} \la -1$; continuous injection models \citep{Pa70} suggest 
that the source age is more than 0.1~Gyr.
Spectral aging in the relic phase, with pitch-angle scattering, will cause a much steeper 
exponential cutoff beyond a break frequency \citep{Ja74}.  The radio colors of the bright rim
along the southern lobe and the NW parts of the northern lobe are consistent with a model in which
the lobes were injected with relativistic plasma with spectrum of index $\alpha = -0.5$ that
subsequently aged. In this model the NW and SW ends of the source correspond to plasma with
relic-phase lifetimes of 40~Myr; the relatively less bright southern rim has a relic life of 55~Myr.
The parts of the two lobes closer to the core have a large scatter in color-color space; however,
their path traces ages from 60--90~Myr with plasma of larger spectral age lying closer to the core.
All these estimates suggest that the activity commenced more than 0.1~Gyr ago and continued until at
least 40~Myr ago.

The NE component of the northern lobe is distinctive in that 
it has $\alpha^{36}_{22} \approx -1$ and significant spectral steepening at higher frequencies; it
has maximum spectral curvature and does not fit any simple model of spectral aging. 
This anomaly, together with the large spread in colors in the 
low-surface-brightness parts of the lobes,
indicates complexity in source history and particle acceleration: the initial spectra created at the ends of the lobes
might have been curved or may have had very different injection indices at different times.
Additionally, all of the above spectral age estimates are likely to be overestimates because 
of expansion out of the hotspots and within the lobes, and because the 
equipartition magnetic fields are smaller in the NE component and 
closer to the core where the surface brightness
is lower and this would shift any spectral break in the electron energy spectrum 
to lower emission frequencies. The lack of inversion symmetry in the color distribution 
properties is additional reason for caution in deriving ages from the color distribution over 
the source.  The morphology itself might in fact be a better indicator of dynamical history.

\subsection{Activity and interaction at the host galaxy}

The host galaxy that is believed to have created the relic radio lobes of SGRS~J0515$-$8100
has a radio power of $10^{22.7}$~W~Hz$^{-1}$ at 5~GHz with a flat non-thermal
radio spectrum indicating ongoing nuclear activity.
Only 10\% of early type galaxies with absolute magnitude corresponding to that
of the host of SGRS~J0515$-$8100 are expected to host AGNs with this radio power \citep{Sa89}
implying that the current activity might be related
to that which created the large-scale radio structures.  An active radio galaxy with total
radio power of $10^{25.3}$~W~Hz$^{-1}$ at 0.408~GHz typically has a core power of 
$10^{23.1}$~W~Hz$^{-1}$ at 5~GHz \citep{Gi88}; SGRS~J0515$-$8100 has a core 
power just a factor of two below this value. Within the scatter in the relationship this is
consistent with the hypothesis that although jets from the core may not be continuing to
supply energy to the lobes, the activity at the core continues at the present epoch. Further
support for this picture come from the relatively small spectral age and 
the edge-brightened structures that suggest activity in the recent past.  

As seen in Fig.~9, the host galaxy appears to be interacting with a fainter 
companion.  The host elliptical is 2.7 magnitudes brighter than the companion and 
both have similar colors suggesting that the mass ratio of the interacting pair is about 12.  
The outer isophotes of the host are offset with respect to the centre of the galaxy and 
this halo is extended away from the companion and towards the far side. The halo probably 
represents a dynamical friction wake --- a relic expected from an encounter with a relatively 
small mass companion \citep{Ba92}.  The companion is observed to have a fine tidal bridge
extending towards the host galaxy and a fine tail on the far side; both features form part
of an arc and these structures are presumably the result of a tidal stretching of the companion
along its post-encounter trajectory.  Such leading and trailing debris trails are expected 
in `victims' after pericenter passage in moderately high speed encounters with galaxies
that are much more massive (see, e.g., \citet{Jo96} and references therein).  The companion is
observed at a projected distance of 80~kpc from the host center.  Assuming that the companion
has an orbital speed that is $\sqrt{2}$ greater than its observed line-of sight velocity, we
estimate that the pair is currently being observed about 0.2~Gyr after pericenter passage.
The data are consistent with a picture where the companion had a close and fast encounter on
a highly elliptic or perhaps hyperbolic trajectory and is now in the foreground of the host, 
moving NE and towards us with a speed about 500~km~s$^{-1}$ relative to the host.  
The blue-shifted emission line gas that is observed towards the host might represent 
foreground gas from the companion infalling towards the host.

\section{The evolutionary history of SGRS~J0515$-$8100}

The model for edge-brightened lobes is one in which
jets terminate in shocks at the ends of the source
and advance the heads at speeds limited by ram pressure of the external gas; this advance
speed is expected to well exceed the lateral expansion speed that is driven by the excess 
pressure in the lobes relative to the medium.  
In comparison with the lobes of powerful edge-brightened radio sources, as well as the relaxed
lobes of Fat-Doubles, the southern lobe of SGRS~J0515$-$8100 is unusual in that 
the lobe is fan shaped with the lobe width increasing towards its end.  This peculiarity in the
morphology, together with the 
extremely small ($< 2$) value of the axial ratio and the edge brightening of the lobes,
suggests that the wide lobes in SGRS~J0515$-$8100 have attained their properties not as a result of
expansion in a relic phase but
as a result of significant (about $50\degr$) variations in the direction of the jet axis
during the history of its activity.  

The evolution in the lobes of SGRS~J0515$-$8100 may have proceeded in two phases: 
(i) an active phase during which energy
was transported to the lobes via jets whose axis varied significantly, and (ii) a recent
relic phase during which the jets were off.  Internal relaxation, continuous
expansion against the ambient intergalactic medium and radiative losses evolved the structure
and spectrum.  The active phase might have been punctuated by
times when the jets ceased and subsequently restarted.
The constraints on the dynamical evolution indicate that if the jet axis were stable over the
source lifetime, and the wide lobes were a result of lateral expansion,
timescales in the range 0.2--4 Gyr are involved in relaxation processes within the lobes
and expansion against the external gas: this exceeds the time since pericenter passage
of the companion.   However spectral aging considerations
suggest timescales of 0.1~Gyr and smaller for the lifetime of the radiating electrons
throughout the source, indicating that what we observe are recently accelerated
electrons that have rapidly streamed across the giant radio lobes.  These radiating electrons throughout
the source were accelerated post pericenter passage of the companion.

A possible scenario is that the radio source lifetime is $\la 0.2$~Gyr
and was triggered by the pericenter passage of the companion.  In this picture the recent close
encounter would also be the cause for the jets to swing around and deposit synchrotron
plasma over a wide angle and, therefore, the limits to lateral expansion speeds due to
ram pressure would not be a constraint on the source age.  However, it has been argued that 
the timescale for a merger event to fuel and trigger jet activity in a massive black hole 
at a galactic center could well exceed 0.1~Gyr (see, for example, \citet{Mo03}).  Additionally, 
the jets in SGRS~J0515$-$8100 have probably switched off about 40~Myr ago, so this constrains the entire 
active life of the giant radio source to be $< 60$~Myr following the galaxy-galaxy interaction.

A more likely scenario is one in which the the activity in the host galaxy
commenced $\ga 1$~Gyr ago with the initial fuel provided by a previous
interaction, perhaps an earlier close passage of the companion itself. 
Subsequent activity over the long dynamical 
timescale led to the formation of the giant radio structure.  The more recent close
encounter with the companion, 0.2~Gyr ago, might have restarted the activity if it had
ceased and, additionally, caused the black hole spin axis and inner accretion disks to
change direction.  

Flyby encounters with low-mass companions that penetrate the outer regions of galaxy halos, 
as is the situation in SGRS~J0515$-$8100, might perturb the gravitational potential deep within the 
half-mass radius and warp disks embedded within AGN hosts \citep{We98,Ve00}.  Such tidal encounters
are believed to be capable of triggering nuclear activity \citep{Li04}, implying that early stages
of encounters could perturb the innermost regions of galaxies.  Additionally, material added in an 
encounter might have angular momentum vector inclined to the accretion disk axis causing warps in
outer parts of accretion disks. 
The flywheel that determines the stability of nuclear jets may not be the angular momentum of the
central black hole, but the more massive accretion disk \citep{Na98} and if flybys cause warps in the
outer accretion disk, the black hole spin axis might realign in 0.1--1~Myr, which is much less 
than the timescale for radio activity.  Because the black hole spin axis is coupled to the outer
accretion disk, it is not implausible that an encounter, which perturbs/warps the disk, might realign
the jet axis.

The relatively short spectral timescales suggest that recent jet activity, 40-100~Myr 
ago, injected the radiating particles whose emission we observe today.  These new jets
were likely directed towards the NE parts of the northern lobe, which are edge brightened,
and towards the SW parts of the edge brightened southern lobe, which has a 
lower fractional polarization.  These particles, produced in the most recent activity phase,
light up the entire radio source by streaming across the lobes and energizing the more
relaxed SW parts of the northern lobe and the eastern parts of the southern lobe where the
fractional polarization is high and the magnetic field presents a relaxed ordered appearance.
The large DR observed might be entrainment related to the galaxy encounter and the 
addition of cold gas to the central parts of the host galaxy.  The scenario sketched here
is in some ways similar to the case of 3C293 (\citet{Ev99} and references therein), 
where past activity triggered by interaction with a gas-rich galaxy 
resulted in extended classical double radio lobes and a recent interaction
with a companion has triggered the formation of a kpc-sized inner double; the renewed activity
in that case was along an axis $30\degr$ offset with respect to the outer older axis.

The observations of SGRS~J0515$-$8100 presented here support the hypothesis that encounters with
significantly lower mass galaxies, which presumably result in minor mergers, might perturb the
inner accretion disk and black hole spin axis direction.  The perturbation manifests as a 
significant change in the direction of jets from the active galactic nucleus.  \citet{Ek02} argue 
that minor mergers, leading to black hole coalescence, could significantly perturb the jet axis
direction and result in X-shaped radio sources: the abundance 
of such radio sources was used to infer the
black-hole coalescence rate and event-rate expectations for gravitational wave detectors.  
The observations of SGRS~J0515$-$8100 indicate that interactions with companions might 
significantly perturb the radio axis direction 
well before the minor merger and any black hole coalescence; 
therefore, evidence for significant changes in jet axis direction does not necessarily imply 
black-hole coalescence.  Consequently, gravitational wave 
event rates inferred from the abundances of such radio sources might be overestimates.

\acknowledgments

The Australia Telescope Compact Array is part of the Australia Telescope
which is funded by the Commonwealth of Australia for
operation as a National Facility managed by CSIRO.  
The MOST is operated by the University of Sydney and supported in part by grants from 
the Australian Research Council.
We acknowledge the use of SuperCOSMOS, an advanced photographic plate digitizing machine 
at the Royal 
Observatory of Edinburgh, in the use of digitized images for the radio-optical overlays.
We thank Kinwah Wu, Garret Cotter, Helen Buttery, Helen Johnston and Shakti
Menon for optical observations and analysis.

\clearpage

\begin{deluxetable}{lll}
\tablewidth{0pc}
\tablecolumns{3}
\tablecaption{Journal of observations}
\tablehead{ \colhead{Band} & \colhead{Telescope} & \colhead{Date} }
\startdata
Radio 36~cm continuum & MOST & 1999~Aug \\
Radio 22 \& 13~cm continuum & ATCA & 2000~Jan, Feb, Jul \& Oct \\
Optical R \& V imaging & ANU 2.3-m & 2002~Jan \\
Optical spectra & ANU 2.3~m & 2002~Jan \& 2004~Jan \\
\enddata
\end{deluxetable}

\clearpage

\begin{deluxetable}{rrrrrr}
\tablewidth{0pc}
\tablecolumns{6}
\tablecaption{Radio flux density measurements of SGRS~J0515$-$8100. \label{tab1}}
\tablehead{
\colhead{Wavelength} & \colhead{Tel/Survey} & 
\colhead{N lobe (mJy)} & \colhead{S lobe (mJy)} &
\colhead{Core (mJy)} & \colhead{Total (mJy)} }
\startdata
 36~cm & MOST & 83 & 210 & --- & 293 \\
 22~cm & ATCA & 50 & 124 & 1.9 & 176 \\
 12~cm & ATCA & 26 & 63 & 2.1 & 91 \\
 6~cm  & PMN  & 13 & 35 & --- & 48 \\
\enddata
\end{deluxetable}

\clearpage

\begin{deluxetable}{crrrll}
\tablewidth{0pc}
\tablecolumns{6}
\tablecaption{Photometry\tablenotemark{a}~~ of the field of SGRS~J0515$-$8100. \label{tab1}}
\tablehead{
\colhead{Object} & \colhead{$v$} & 
\colhead{$r$} & \colhead{$V-R$} &
\colhead{Class} & \colhead{Ref\tablenotemark{b}} }
\startdata
a & 20.13 & 19.68 & 0.47 & K0 V & \citet{Be90} \\
b & 16.89 & 16.20 & 0.71 & E(z=0.1) & \citet{Fu95} \\
c & 21.19 & 20.76 & 0.46 & K0 V & \citet{Be90} \\
d & 20.28 & 19.38 & 0.84 & K7 V & \citet{Be90} \\
e & 19.56 & 18.88 & 0.71 & E(z=0.1) & \citet{Fu95} \\
\enddata
\tablenotetext{a}{$v$ and $r$ are the observed magnitudes, while $V$ and $R$ are
magnitudes in the standard Johnson-Morgan/Cousins color bands.}
\tablenotetext{b}{References for the classification.}
\end{deluxetable}

\clearpage
\begin{figure}
\epsscale{0.7}
\plotone{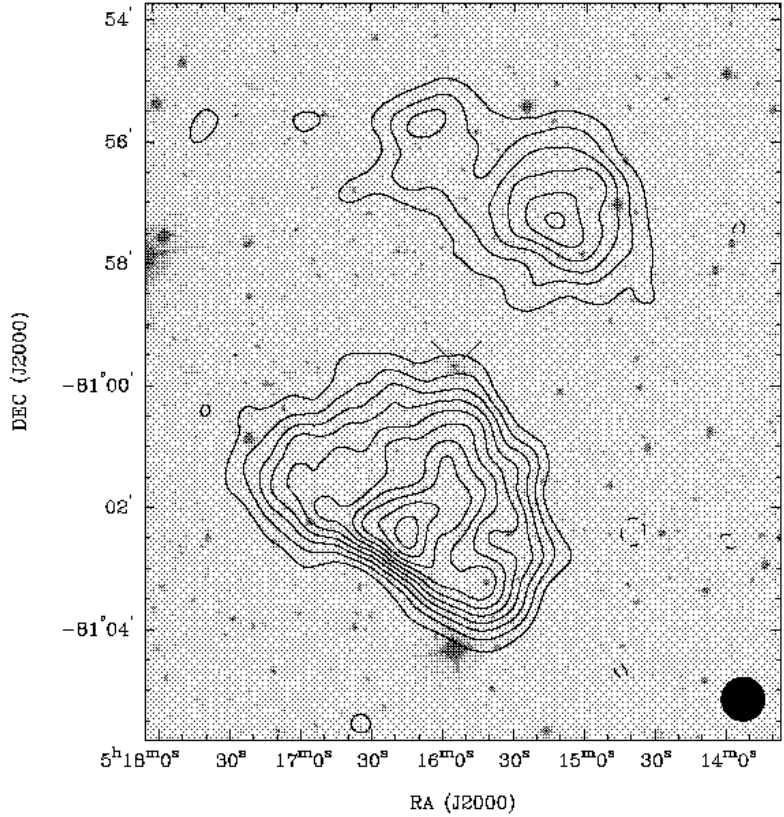}
\caption{ Contours of the 36-cm MOST image of SGRS~J0515$-$8100 
overlaid on a grey scale representation of the SuperCOSMOS digitization
of the UKST blue image of the field.  The radio image has a beam of FWHM 
$43\farcs5 \times 43\farcs0$ at a P.A. of 0$\degr$. Contours are 
at $-$1.5, 1.5, 3, 4.5, 6, 7.5, 9, 10.5, 12, 13.5 and 15 mJy~beam$^{-1}$; 
the lowest contour is at a level of 3 times the rms noise in the image. 
The host galaxy at the center of the image is indicated by a pair of thick lines.
In this image, as well as all following images displayed herein, the 
half-maximum size of the beams of the radio images are shown using a filled ellipse 
in the lower right of the figure.  Additionally, all the radio images have been
corrected for the attenuation due to the primary beam.
\label{fig1}
	}
\end{figure}

\clearpage
\begin{figure}
\epsscale{0.7}
\plotone{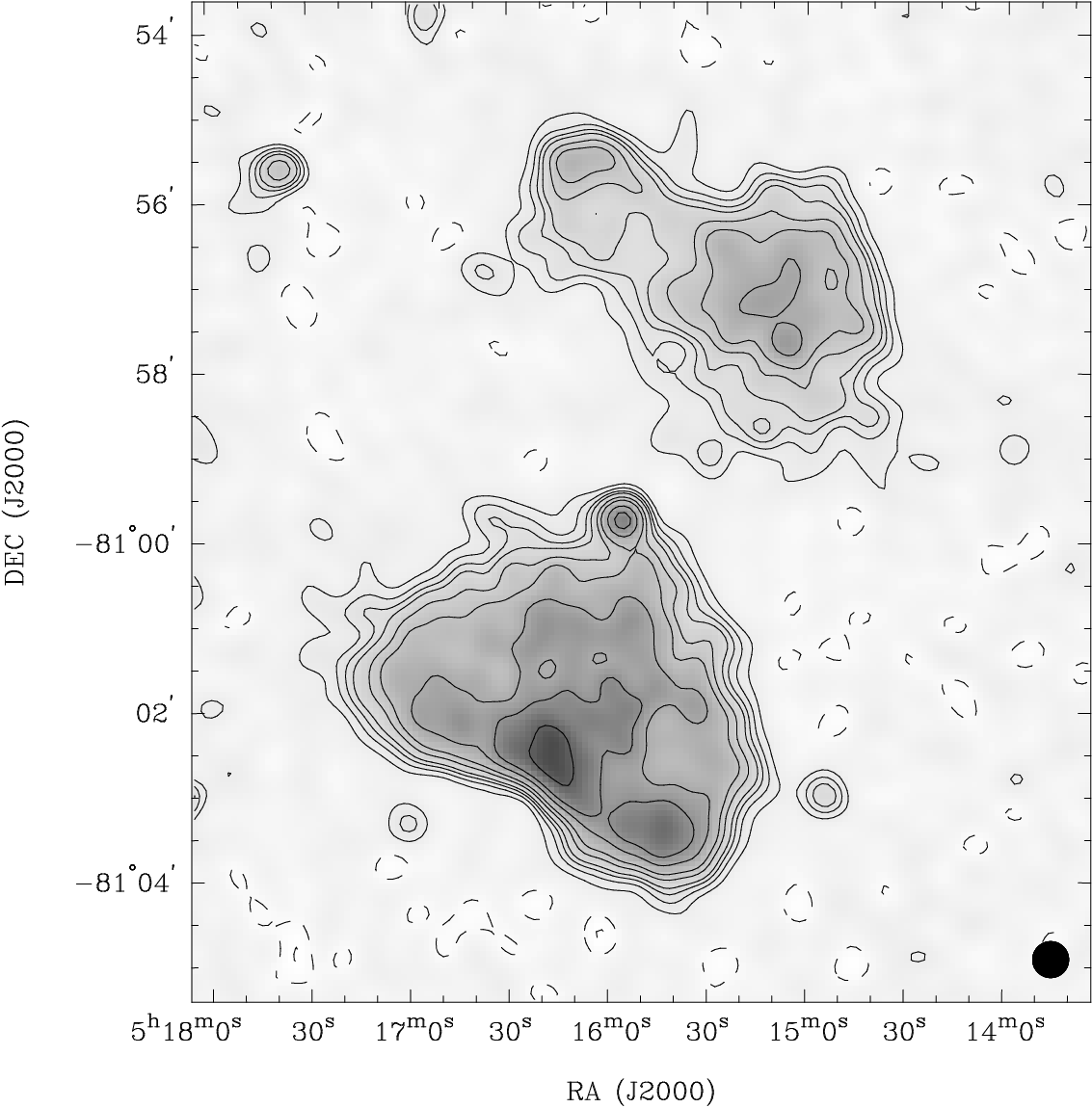}
\caption{22~cm image of SGRS~J0515$-$8100 made using the ATCA with a beam of
$25\arcsec$ FWHM. Contours are shown at ($-2$, $-1$, 1, 2, 3, 4, 6, 8, 12, 16 and 24) $\times$
0.15 mJy~beam$^{-1}$, the lowest contour is at 3 times the rms noise.  
Grey scales are linear and span the range $-$0.3 to 6 mJy~beam$^{-1}$.  
\label{fig2}
	}
\end{figure}

\clearpage
\begin{figure}
\epsscale{0.7}
\plotone{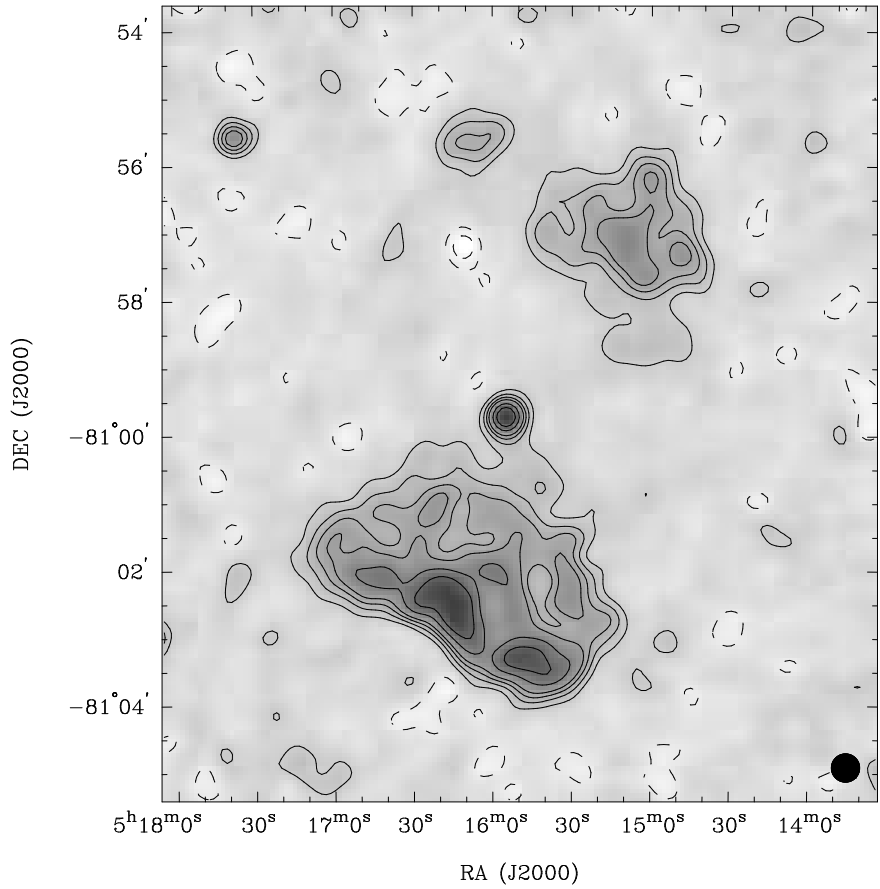}
\caption{12~cm image of SGRS~J0515$-$8100 made using the ATCA with a beam of
$25\arcsec$ FWHM. Contours are shown at ($-2$, $-1$, 1, 2, 3, 4, 6 and 8) $\times$
0.2 mJy~beam$^{-1}$, the lowest contour is at 2.5 times the rms noise.  
Grey scales are linear and span the range $-$0.5 to 3 mJy~beam$^{-1}$.  
\label{fig3}
	}
\end{figure}

\clearpage
\begin{figure}
\epsscale{0.7}
\plotone{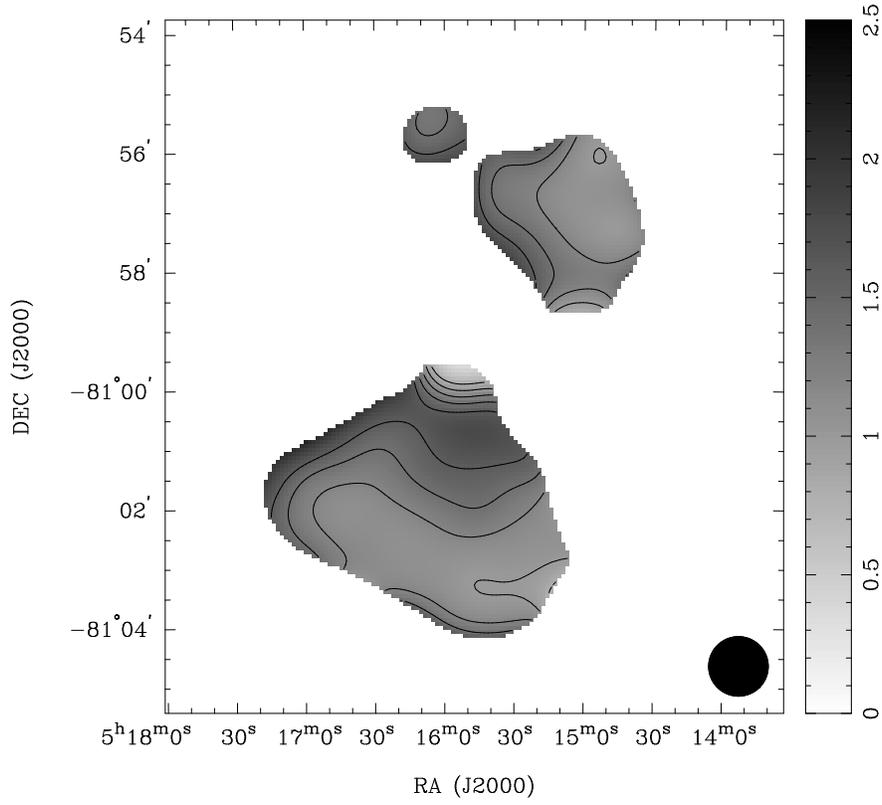}
\caption{The distribution in the radio spectral index over SGRS~J0515$-$8100;  $-\alpha$, computed 
between 36 and 12~cm using images made with beams $1\arcmin$ FWHM, is shown using grey scales 
in the range 0--2.5 and contours at 0.8, 1.0, 1.2, 1.4 and 1.6.
\label{fig4}
}
\end{figure}

\clearpage
\begin{figure}
\epsscale{0.7}
\plotone{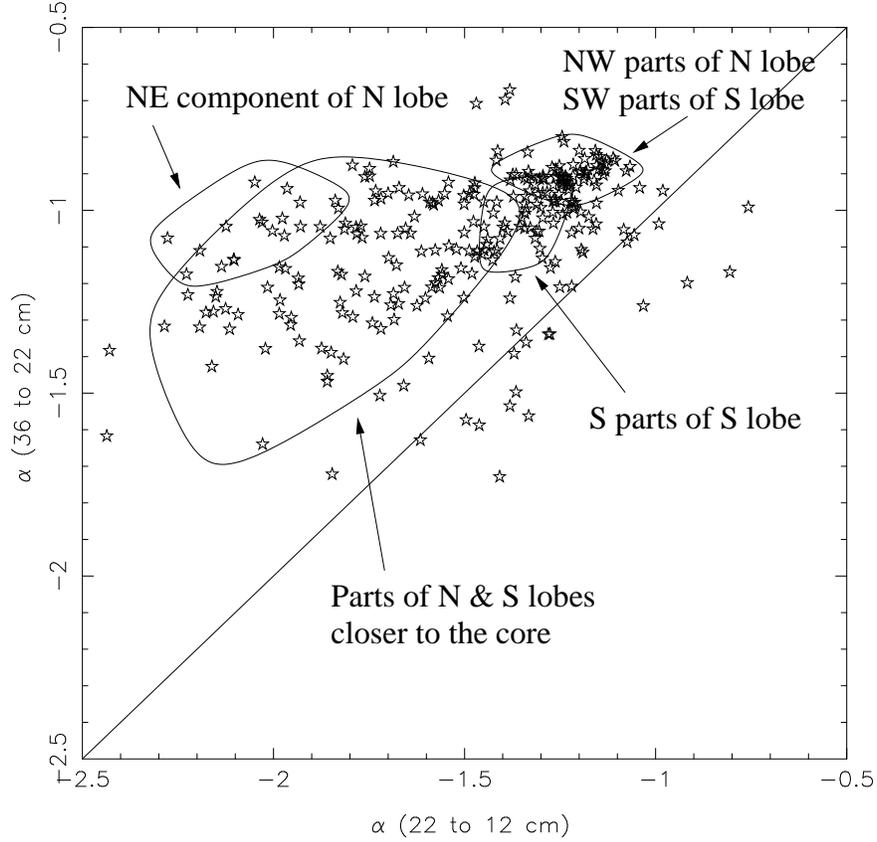}
\caption{A color-color plot.  Radio continuum images 
of SGRS~J0515$-$8100 were made with beams $1\arcmin$ FWHM at 36, 22 and 12~cm. 
The pixel intensities
were used to compute spectral indices $\alpha^{36}_{22}$ between 36 and 22~cm 
and $\alpha^{22}_{12}$ between 22 and 12~cm.  The figure
shows the distribution of image pixels over the plane with $\alpha^{22}_{12}$ along the x axis 
and $\alpha^{36}_{22}$ along the y axis. Pixels over the core component have been omitted.
The diagonal line represents the locus of image pixels that have
a single power-law spectrum over this wavelength range.
\label{fig5}
}
\end{figure}

\clearpage
\begin{figure}
\epsscale{0.7}
\plotone{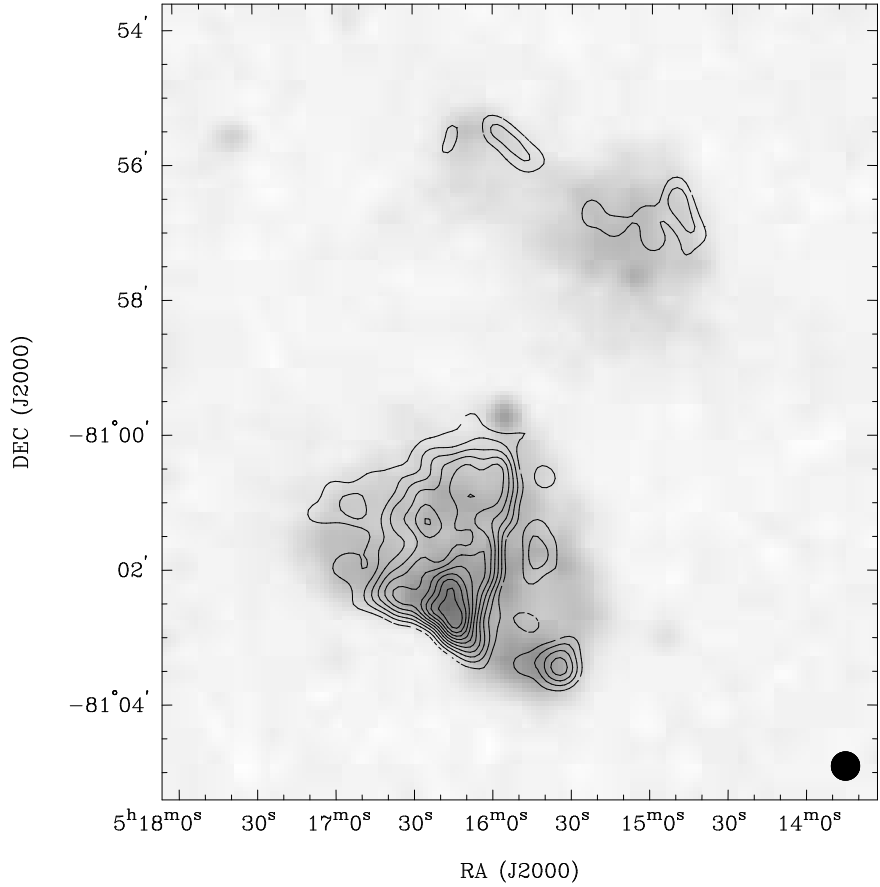}
\caption{Contours of the distribution in the intensity of the 22-cm linear polarization overlaid
on a grey scale representation of the 22-cm total intensity image.  Contours are at 0.3, 0.45, 0.6,
0.75, 0.9, 1.05, 1.2, 1.35, 1.5 and 1.65~mJy~beam$^{-1}$; grey scales cover the range $-0.4$
to 8.0~mJy~beam$^{-1}$. Both images have beams of FWHM $25\arcsec$.
\label{fig6}
}
\end{figure}

\clearpage
\begin{figure}
\epsscale{0.7}
\plotone{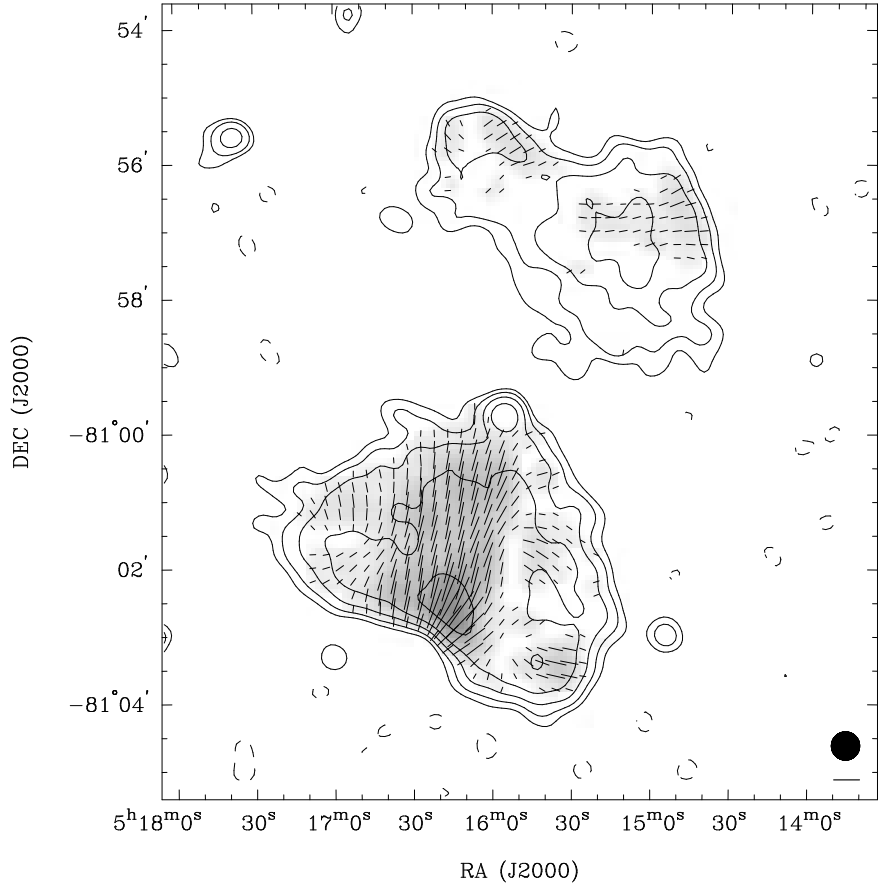}
\caption{The distribution in the projected E field corrected for Faraday rotation.  
Vector lengths represent the polarized intensity and orientations the P.A. of the E field.  
Contours of the
total intensity at $-0.2$, 0.2, 0.4, 0.8, 1.6 and 3.2~mJy~beam$^{-1}$, grey scales
represent the polarized intensity over the range 0--4~mJy~beam$^{-1}$.  The scale bar
below the beam FWHM circle corresponds to a polarized intensity of 1~mJy~beam$^{-1}$. 
All images have beams of FWHM $25\arcsec$.
\label{fig7}
}
\end{figure}

\clearpage
\begin{figure}
\epsscale{0.7}
\includegraphics[angle=-90]{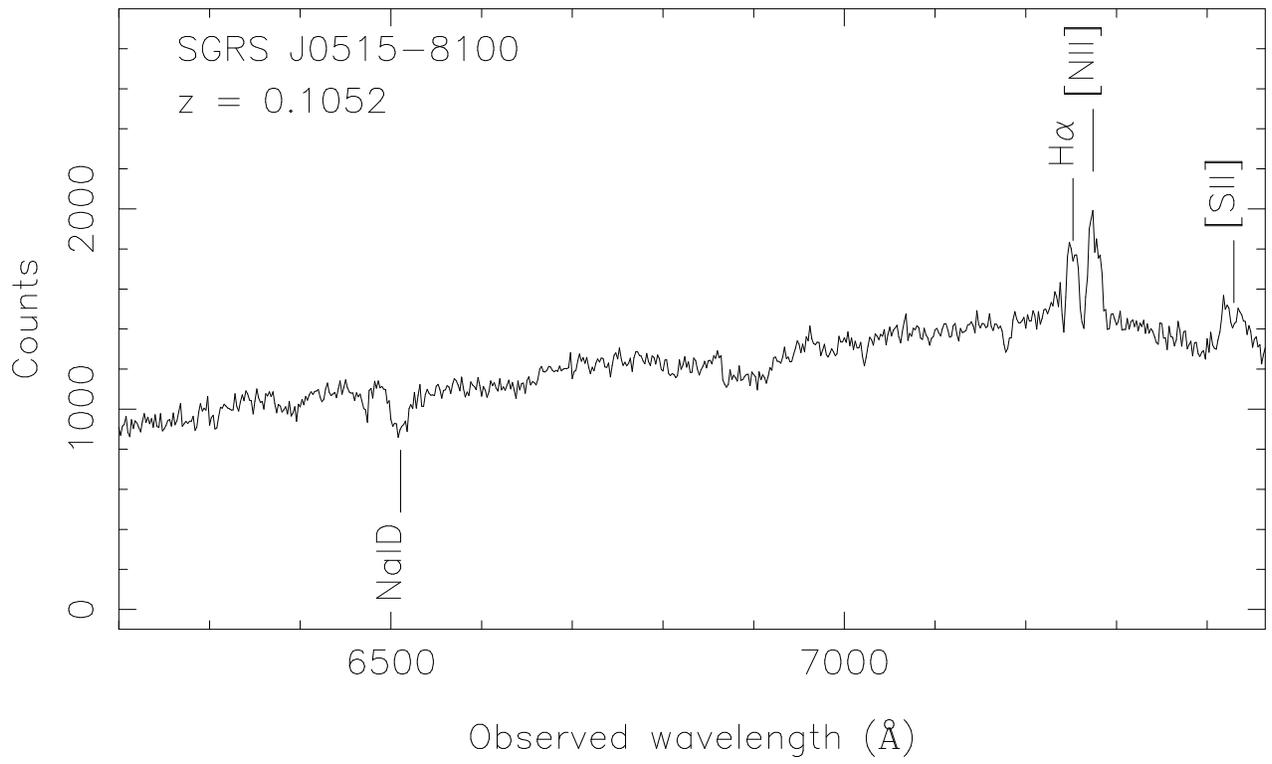}
\caption{Optical spectrum of the host galaxy in the red.  The H$\alpha$ and [N {\sc ii}] lines
have velocity widths about 200~km~s$^{-1}$ and signs of velocity structure within the lines.
\label{fig8}
}
\end{figure}

\clearpage
\begin{figure}
\epsscale{1.05}
\plotone{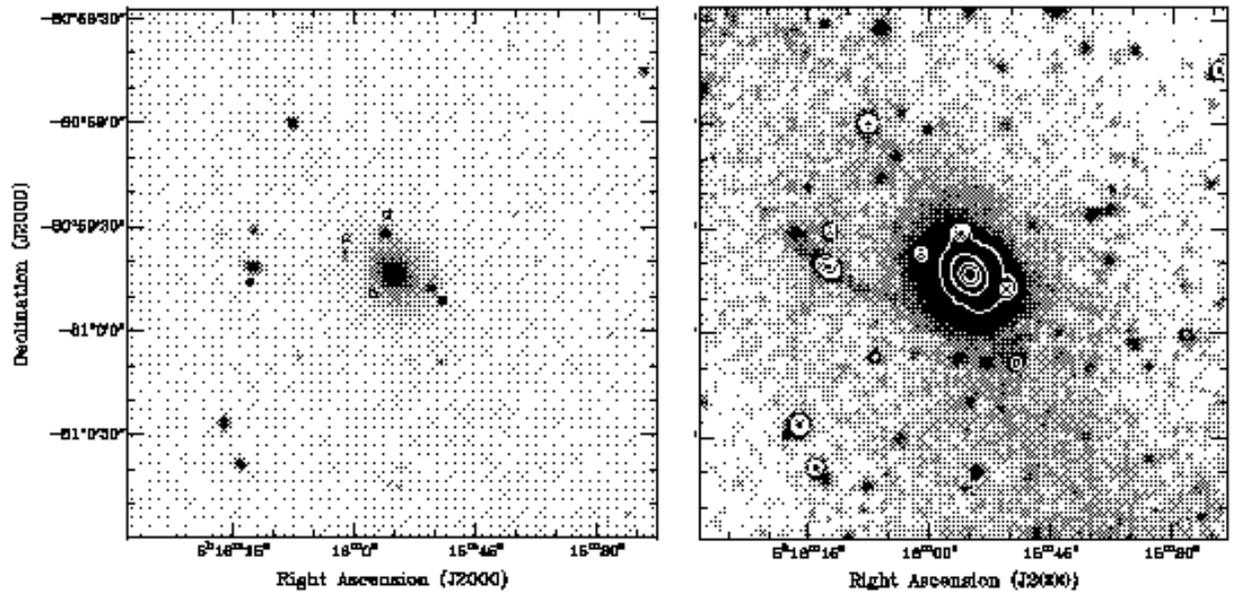}
\caption{Images of the optical field in the vicinity of the host galaxy. Objects in the
field are labelled in the panel on the left. The host is the object marked `b' and `e' 
is a companion with signs
of tidal interaction; the grey scales have been adjusted in the panel on the
right to display the tidal tails, with contours showing the structure obliterated in this
high contrast display.
\label{fig9}
}
\end{figure}

\clearpage
\begin{figure}
\epsscale{0.7}
\includegraphics[angle=-90]{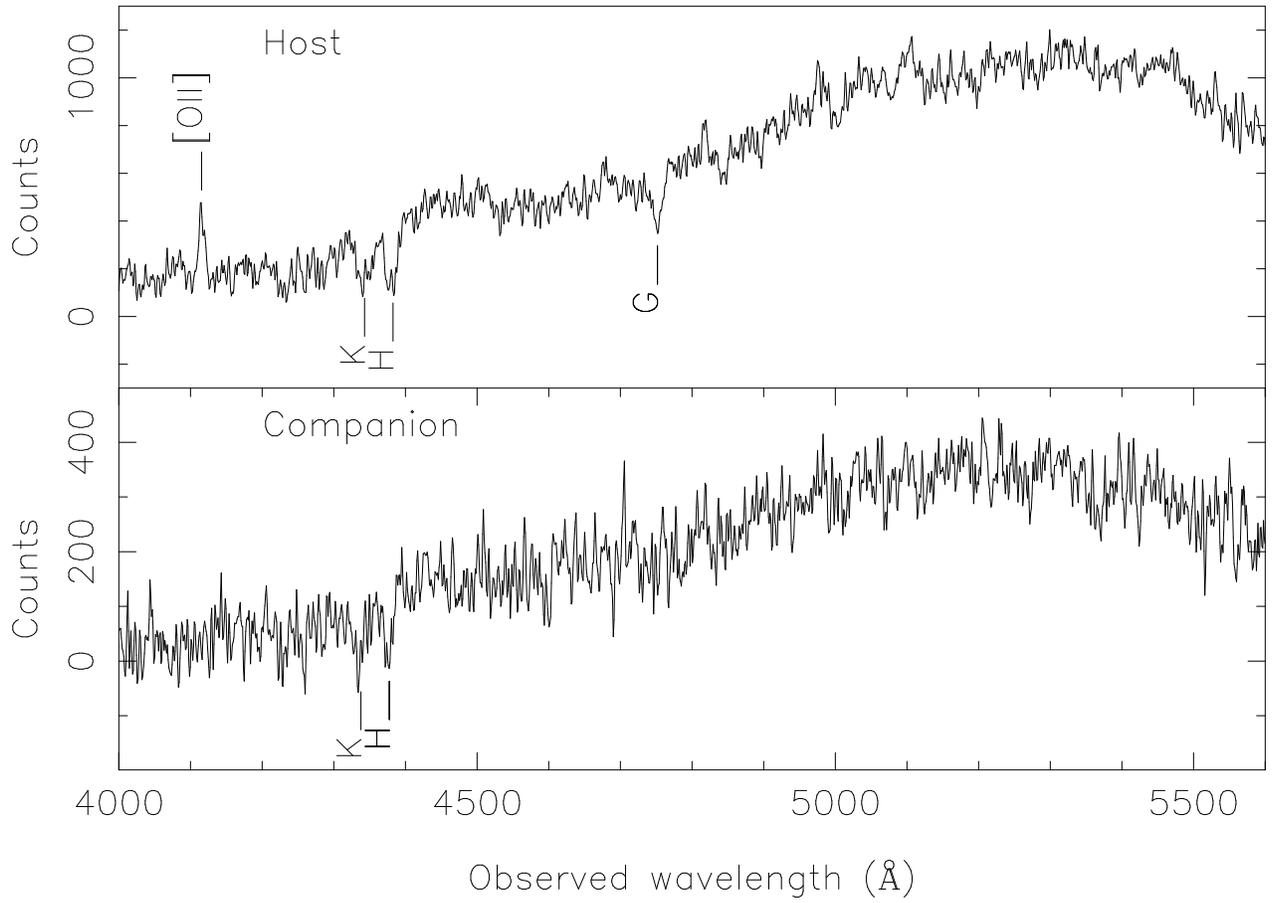}
\caption{Optical blue-region spectra of the host galaxy (upper panel) and its 
companion (lower panel).  
\label{fig10}
}
\end{figure}

\end{document}